\pdfoutput=1

\documentclass[final,1p]{elsarticle}

\usepackage[utf8]{inputenc}
\usepackage[american]{babel}

\usepackage{color}
\usepackage{multirow}
\usepackage{array}
\usepackage{listings}

\usepackage{amsfonts}
\usepackage{amsmath}

\usepackage{graphicx}

\usepackage[lined,boxed]{algorithm2e}

\date{}

\journal{Computer Networks}

\begin{document}

\begin{frontmatter}

\title{Consistency Maintenance of State of Management Data in P2P-based Autonomic Network Management}

\author{Jéferson Campos Nobre\corref{cor}}
\cortext[cor]{Corresponding author.}
\ead{jcnobre@inf.ufrgs.br}
\author{Lisandro Zambenedetti Granville}
\ead{granville@inf.ufrgs.br}

\address{Institute of Informatics - Federal University of Rio Grande do Sul \\
      Av. Bento Gon\c{c}alves, 9500 - Porto Alegre, RS - Brazil}

\begin{abstract}

Complex Dynamic Networks can be exploited in solving problems where traditional solutions may not be sufficient. 
The increasing complexity of computer networks imposes problems to the current network management solutions. 
In this context, network management is an example of a research area that could benefit from the use of CDNs. However, the consistency of state of management data among the elements that build management CDNs (management nodes) is an important challenge. 
Traditional mechanisms to maintain consistency of these states are supported by some centralization which wastes some desirable properties of CDNs (\textit{e.g.}, robustness). 
In contrast to these mechanisms, we propose a distributed, scalable and robust mechanism to maintain the consistency of state of management data in management CDNs. 
Our mechanism introduces multi-agent truth maintenance features and communication strategies based on dynamic process to provide consistency maintenance of state of management data. We developed a model of a management CDN on Peersim simulator to perform experiments. 
Besides, 2 case studies are presented. The result obtained supports our scalability and robustness claims.

\end{abstract}


\end{frontmatter}

\section{Introduction}



Complex Dynamic Networks (CDNs) are networks whose dynamic and collective behavior is a function of the sophisticated individual properties of nodes and links that compose them \cite{CDN-Strogatz-2001}. CDNs are found and observed in several fields, varying, for example, from biology (\textit{e.g.}, immune systems) to computer science (\textit{e.g.}, social networks). As such, different research initiatives have been investigating how to model and analyze CDNs in order to better understand their behavior \cite{CDN-Newman-2003}. Inspired by studies of real-world CDNs, such initiatives have identified important CDN characteristics, such as network evolution and dynamical complexity \cite{CDN-Strogatz-2001}. From another perspective, CDNs can also be exploited in solving problems where traditional solutions may not be sufficient. The management of computer networks is an example of a research area, which is key to this article, that could benefit from the use of CDNs.


The complexity of computer networks, as well as their use and importance, has increased significantly in recent years. However, computer network advancements are usually not accompanied by corresponding adequate management solutions \cite{ANM-Samaan-2009}. One possibility of tackling this problem is by incorporating CDN characteristics into network management, especially if one considers that these characteristics (\textit{e.g.}, improved fault tolerance, distributed processing, and load balance) are usually expected, but not always found, in today's network management systems. In fact, some management solutions proposed by the network management community (\textit{e.g.}, mobile agents and peer-to-peer for network management) can be described as CDNs themselves. The recent use of CDNs concepts in network management, however, also introduces new problems that need to be addressed.


A CDN for network management is an overlay network running on top of the communication infrastructures that need to be managed. A management CDN is composed of peers that have a double role: besides acting as regular CDN nodes (\textit{i.e.}, performing operations to support the CDN), they also execute management tasks over the underlying managed infrastructure. CDN nodes themselves need to store and exchange data as a part of these tasks. However, the state of management data stored in these nodes can become inconsistent because of, for example, the dynamicity of nodes' interaction. For example, in automated and distributed decision making processes, management decisions must be advertised by different nodes of the management CDN in a coordinated and coherent way. These advertisements are directed towards the interested human administrators or management applications. If incoherent advertisements take place (\textit{i.e.}, some nodes advertise different results for the same decision), they can lead to instabilities in the underlaying network management. Thus, a mechanism to maintain the consistency of state of management data is necessary.


Despite the importance of having consistency mechanisms for the state maintenance of management data in today's management CDNs, this topic has been barely addressed in previous investigations, as will be further discussed in Section \ref{sec: Scenario}. In fact, when present, such mechanisms are in general materialized in centralized solutions \cite{AUTONOMIC_POLICIES-Badr-2004} \cite{ANM-Fallon-2007}, which hinder distribution benefits (\textit{e.g.}, scalability), in addition of being essentially inconsistent with the distributed nature of CDNs. Even when distribution indeed takes place, these mechanisms are usually not clearly described in the distributed management literature \cite{ANM-Jennings-2007} \cite{AC_NM-Marquezan-2007}.


In this article we present a solution for the maintenance of state consistency of management data in management CDNs. The contribution of this article is twofold. First, we propose the use of \textit{belief exchange} about management data to improve the consistency of states of management data in a management CDN. The proposed belief exchange is inspired by multi-agent truth maintenance \cite{TMS-Huhns-1991}, a concept that we borrow from Multi-Agent Systems (MAS) \cite{MAS-Sycara-1998}, where states of management data are organized as a set of justified beliefs among the management nodes. Second, we present communication strategies in order to support belief exchange among management nodes. These strategies use dynamic bio-inspired processes (proliferation), which are recognized as scalable and robust \cite{BIO-Babaoglu-2006}. The employment of these communication strategies creates multi-layered management CDNs.





The remainder of this article is organized as follows. In Section \ref{sec: Scenario} we present a motivating scenario, while in Section \ref{sec: Justification} we detail how the concept of belief exchange is adapted to consistency maintenance of management data in the context of management CDNs. We describe the architecture of a management node in Section \ref{sec: Architecture} and we show case studies in Section \ref{sec: Case Study}. We introduce the communication strategies for the belief exchange among management peers in Section \ref{sec: Communication}. In Section \ref{sec: Evaluation} we evaluate our proposal discussing the results obtained with simulation experiments. We compare our research with related work in Section \ref{sec: Related Work} and finally close this article with concluding remarks and future work in Section \ref{sec: Conclusion}.

\section{State consistency for management CDNs: P2P-based autonomic network management as a motivating scenario}
\label{sec: Scenario}


There has been substantial research on autonomic features related to network management solutions \cite{ANM-Samaan-2009}. The application of Autonomic Computing (AC) principles in network management, normally referred as \emph{Autonomic Network Management} (ANM), has been proposed as a way to address some demands faced by traditional network management, such as controlling highly dynamic environments as in \textit{ad hoc} networks \cite{NM-Pras-2007}. Several authors claim that some level of decentralization plays an important role to perform autonomic actions in a more adequate manner (\textit{e.g.}, \cite{AUTONOMIC-Dobson-2006}). Different technologies could be employed as an infrastructure of a decentralized ANM system. An interesting possibility is using P2P overlays, which incorporates characteristics of P2P-based network management (P2PBNM) into ANM systems, such as the robustness in connectivity of management entities \cite{NM_P2P-Granville-2005}.

The constitutive peers of a P2P-based ANM system interact dynamically and behave collectively, given their individual dynamics and coupling pattern. Thus, they can be defined as management nodes of a management CDN. These nodes expose dynamic properties because of their autonomy capabilities in the execution of management tasks. Other examples of systems taking the form of CDNs abound in the world, such as, social networks, networks of business relations between companies, metabolic networks, food webs, blood vessels, power grids, and networks of citations between papers \cite{CDN-Strogatz-2001} \cite{CDN-Newman-2003}.


The execution of management tasks can change the state of management data among nodes. However, this state must be consistent across a management overlay. Thus, consistency maintenance procedures play an important role in management CDNs. These procedures must handle heterogeneity in these CDNs. Management nodes can be heterogeneous, being deployed in management stations or embedded into network devices. Besides, management tasks can be also heterogeneous, being requested by human network administrators or triggered by state-of-the-art Machine-to-Machine (M2M) interactions \cite{M2M-Lawton-2004}. A biological analogy that can be used is the coherent and stable behavior from cells that form tissues of complex multicellular organisms.


In this Section, we present an overview of a typical consistency maintenance procedure in a P2P-based ANM system. Afterwards, the most significant P2P-based ANM intitiatives found in the literature and their consistency maintenance mechanisms are described. 

\subsection{A typical consistency maintenance procedure in a P2P-based ANM system}

Before introducing the issues associated to a typical consistency maintenance procedure, consider the environment of a decentralized ANM system. It is often composed of a set of Autonomic Management Elements (AME), and these AMEs are often grouped into Autonomic Management Domains (AMD). In P2P-based ANM initiatives, peers have some properties found in AMEs and peer groups have some properties found in Autonomic Management Domains (AMD). The interactions among AMEs of an AMD can be characterized as a management CDN.


P2P-based ANM systems use the P2P overlay communication services to propagate messages among the peers. This overlay can be modeled as an asynchronous unreliable communication network, thus it must be considered some issues, such as message delay and loss. In this context, P2P-based ANM systems can be described as asynchronous distributed systems. Besides, it is well known that the utilization of asynchronous distributed systems imposes challenges to achieving consensus or agreements \cite{FT-Fischer-1986}. Thus, the consistency maintenance mechanism aims in overcome these challenges to provide support for the consistency maintenance of states of management data. To illustrate a consistency maintenance procedure, it is described a change in the state of a specific management datum due the reception of an asynchronous signal by a decentralized ANM system. In this example, this change - a new state of a management datum - must be consistently stored in the system's Knowledge Base (KB).
 

The traditional consistency maintenance mechanism is the utilization of a shared repository for storing states of management data. Thus, state of management data is  centrally-stored, whether in a special peer or even in a regular server (\textit{i.e.}, any single point of storage). This mechanism can be seen in some P2P applications through the use of a ``central server'' (\textit{e.g.}, Napster). The consistency maintenance procedure performed by this mechanism is defined as follows. After receiving an asynchronous signal, the peers send a message to the server to inform the change in the state of management data. After these messages, the server sends a message to peer group and peers check whether the state is coherent with their KB. If it is not, the peers update their KB.


There are some major issues related to such centralized approach to deploy this consistency maintenance mechanism. First, the utilization of a shared resource presents robustness issues, since the shared repository is a single point of failure (SPoF). As a consequence, this repository is often implemented using clusters. Second, there are robustness issues because the shared repository represents a bottleneck in terms of computation and bandwidth. This may translate into increases in the total cost of ownership (TCO) to support a bigger management infrastructure. Third, this centralization decreases peers' autonomy. This can derail the operation of a fully decentralized ANM system. These issues underscore the utilization of shared repositories to maintain the consistency of states of management data.

\subsection{P2P-based ANM initiatives and their consistency maintenance mechanisms} 


There are several initiatives investigating P2P-based ANM. These initiatives have different consistency maintenance mechanisms. For sake of simplicity, only few initiatives in P2P-based ANM will be cited, as follows.


PBMAN \cite{ANM-Kamienski-2006} merges traditional PBNM with P2P overlays to manage Ambient Networks (AN). PBMAN enables management tasks inside the AN, as well as distribution and retrieval of management policies. Through this approach it is possible to manage seamlessly devices or services. PBMAN is structured using super peers, in a hierarchical architecture. States of management data are not effectively distributed in this initiative, and, for fault tolerant features, super peers are replicated.


ManP2P \cite{NM_P2P-Panisson-2006} is an example of P2P-based network management system that is evolving to an autonomic conception through the implementation of autonomic modules in peers. ManP2P is partially inspired by the Management by Delegation (MbD) model and based on a service-oriented approach. The autonomic modules are designed to support self-* features and to communicate with other components of the ANM system, when necessary. The distribution of states of management data is not clearly described in this initiative \cite{AC_NM-Marquezan-2007}.


Self-Managed Cells (SMC) \cite{ANM-Lupu-2007} are proposed as an architectural pattern for ubiquitous computing applications, aiming at different levels of scale. Each SMC is autonomous and uses policy-based techniques for driving adaptation decisions. Among different cross-SMC interactions, the authors describe P2P interactions. But, concerning to levels of abstraction, it is not clear whether different SMCs could be ``peers''. A managed device is logically connected with only one SMC, thus, the state of management data is not evaluated in a P2P fashion.


The Madeira platform \cite{ANM-Fallon-2007} is an approach to network management topologies that uses the concept of Adaptive Management Components (AMC), which are containers that run on managed elements. AMCs can manage elements on which they are running and communicate with other AMCs running on other managed elements through P2P communication services. AMCs form management clusters, where a cluster head (\textit{i.e.}, super peer) coordinates the cluster and correlates data. The use of cluster heads centralizes the maintenance of state of management data.

Despite many improvements brought by P2P-based ANM systems (\textit{e.g.}, higher availability of network management system), there are still issues to be addressed. States of management data in different peers of an P2P-based ANM system can become inconsistent in overlay operation due to faults or lack of synchronization. Besides, the consistency maintenance of these states must be done maintaining scalability and robustness features of P2P overlays. This consistency maintenance is equally important to any management CDN to present a coherent behavior as a whole. \\


\noindent The scenarios faced by management CDNs (exemplified by P2P-based ANM systems in this article) to maintain consistency of states of management data are similar to those faced in different Multi-Agent Systems (MAS) \cite{MAS-Sycara-1998}. The operation of a weakly coupled infrastructure hampers the utilization of conventional strategies such as explicit message exchange or shared memory for coordination purposes \cite{MAS-Dimopoulos-2006}. In a MAS, agents must be able to assess and maintain the integrity of the information exchanged among them and conflicts about contradictory knowledge may arise during the communication. Besides, agents have to attend their tasks in an asynchronous way. Thus, techniques from MAS could be used to improve the consistency of states of management data in management CDNs.

\section{Justifications for Management Data} \label{sec: Justification}




In a management Complex Dynamic Network (CDN), nodes that perform a specific task must share management data. In this article, \emph{management datum} is defined as a management information described in a defined form (\textit{i.e.}, using a specific language). Management CDNs use several sets of management data in their operation, which can be integrated into \emph{Knowledge Bases} (KBs). The consistency maintenance of these KBs is one of the main challenges related to management CDNs, such as P2P-based ANM systems. In these CDNs, KBs are distributed (and shared) among the elements that build these networks (\textit{i.e.}, management nodes).


Management data must allow their use in distributed automation and/or optimization procedures in a management CDN, which hampers the consistency maintenance of these data. For example, learning and reasoning techniques in distributed management ontologies can be present in these procedures \cite{ANM-Jennings-2007}. It is also expected that sources of management data (\textit{e.g.}, highly dynamic environments) impose challenges to the management CDN. Despite these challenges, it is necessary to avoid potential inconsistencies in the state of management data among nodes. This state among different nodes builds another layer over the management CDN, \textit{i.e.}, creating a multi-layer overlay.


Our proposal is aimed at meeting consistency requirements of state of management data in a management CDN. The proposed mechanism introduces \emph{multi-agent truth maintenance} \cite{TMS-Huhns-1991} features through a \emph{consistency maintenance module} (this module is described in Section \ref{sec: Architecture}) that runs in each node. These features are represented by belief exchange about management data through the utilization of justifications. In this context, management nodes begin to display some characteristics of intelligent agents and the management CDN aggregates some features related to Multi-Agent Systems (MASs). As far as we are aware of, the only studies that incorporate multi-agent truth maintenance features in network management were carried out by Nobre and Granville \cite{ANM_POLICIES-Nobre-2009} \cite{ANM_ETH-Nobre-2009} \cite{ANM-Nobre-2010}.


In this Section, we present the utilization of \emph{justifications for management data}. Initially, we present an overview of truth maintenance in order to provide a theoretical background (Subsection \ref{sec: Truth Maintenance}). Afterwards, our conceptual solution and an internal representation are described (Subsection \ref{sec: Conceptual Solution}). 

\subsection{Truth Maintenance}
\label{sec: Truth Maintenance}

Truth Maintenance Systems (TMSs) were proposed to keep the integrity of KBs. The origin of these systems was proposed in the 1970s, for resolutions in mono-agent systems \cite{TMS-Doyle-1978} \cite{TMS-Doyle-1979}. TMSs provide considerable power using few computational resources \cite{TMS-Kagal-2008}. Although not being well known outside artificial intelligence community, TMSs are used in different contexts, such as policy systems \cite{TMS-Kagal-2008}, recommendation systems \cite{TMS-Lorenzi-2007}, plan adaptation and repair \cite{TMS-Warfield-2007}, and ontology schemes \cite{TMS-Clark-2007}.


TMSs keep the integrity of KBs performing belief revision and exchange in a set of beliefs. These systems keeps track of logical structure of the set of beliefs of agents. A belief is a member of the current set of beliefs if it has valid reasons (\textit{i.e.}, the belief is well-founded). Usually, TMS is implemented through a software module inside an agent. The evaluation of logical consistency of the KB is done by another module, the Problem Solver (PS) \cite{TMS-Huhns-1991}. The PS sends beliefs and their respective foundations to TMS, which, then, registers and associates foundations with respective beliefs. For instance, in policy systems, the PS role is played by the software component that handles policy processing.




TMSs have been extended for MAS versions (\textit{e.g.}, Distributed Truth-Maintenance Systems \cite{TMS-Huhns-1991}). These multi-agent extensions are commonly refereed as \emph{multi-agent truth maintenance}. In a MAS, agents must be able to maintain the integrity of their KBs, despite message exchange with other agents. In order to perform truth maintenance, each agent has its own TMS, thus maintaining the integrity of their KBs.

\begin{figure}[htb]
  \begin{center}
     \includegraphics[scale=.33]{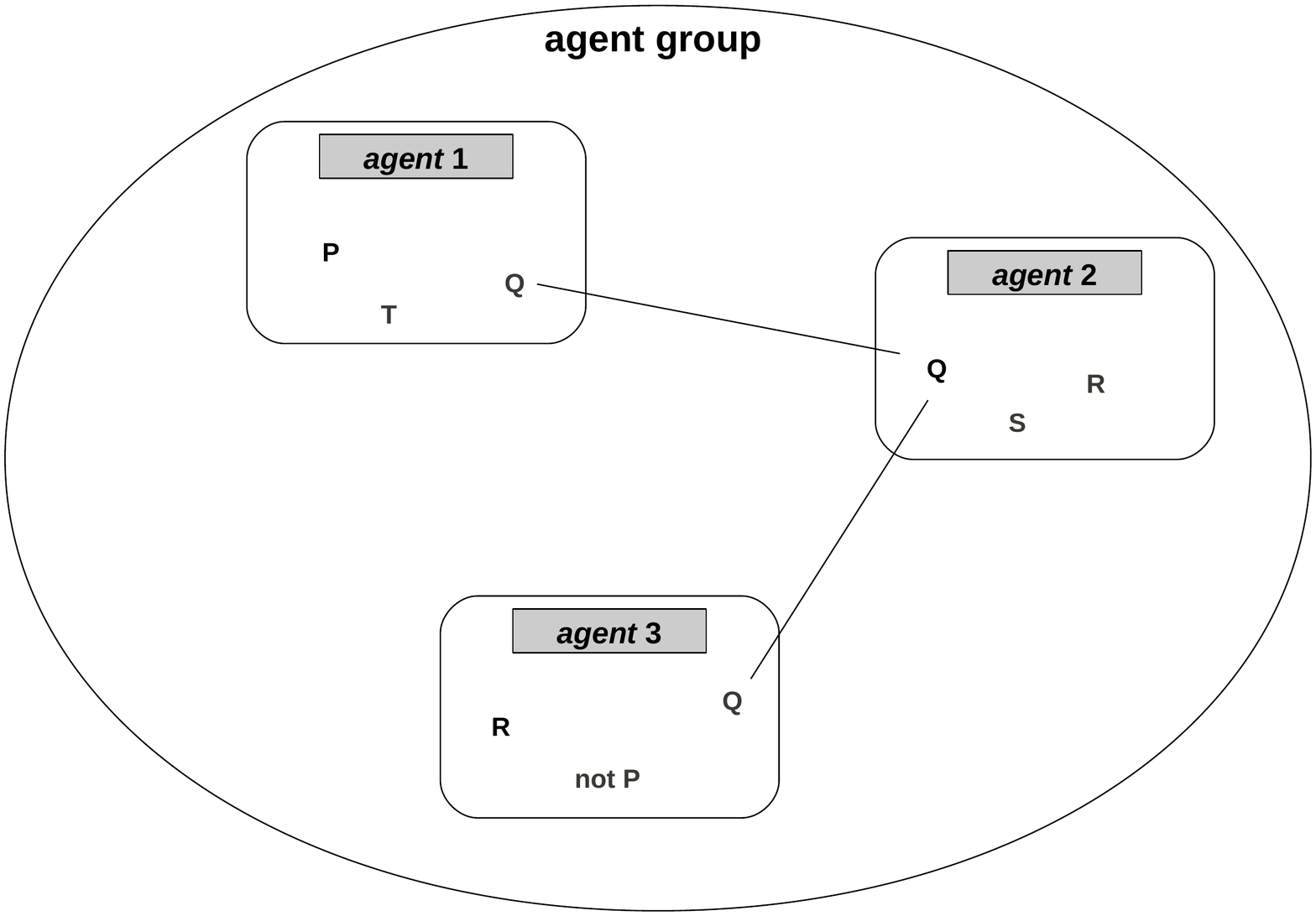}
  \end{center}
  \caption{Shared and Private Data - Figure adapted from \cite{TMS-Huhns-1991}}
  \label{fig:TMS_beliefs}
\end{figure}

Each agent keeps its local data structure. However, agents can be heterogeneous, since agents can have distinct data sets. In this context, data inside agents can be divided in 2 classes: shared data, beliefs that the agent has shared with another agent at some time in the past; and private data, beliefs that the agent has never shared with another agent \cite{TMS-Huhns-1991}. The relation between shared and private data among agents is shown in Figure \ref{fig:TMS_beliefs}.


Figure \ref{fig:TMS_beliefs} describes a group of 3 agents in a given moment. Besides, it is shown some data stored by each node. The ``Q'' datum is the only one shared inside this agent group. The other data represented in this Figure (``R'' and ``T'') are private with respect to specific agents. Agents need to be consistent just about data they all share. Thus, according to the agent group described in Figure \ref{fig:TMS_beliefs}, only beliefs about the ``Q'' datum would be exchanged inside this group.


One of the possible alternatives to implement multi-agent truth maintenance features is the utilization of \emph{justification-based} TMSs \cite{TMS-Huhns-1991}. In a justification-based TMS, a datum is believed when it has valid justifications (\textit{i.e.}, valid reasons). Justifications are produced according to available information, in a defined way (\textit{i.e.}, understood by the TMS) \cite{TMS-Doyle-1979}. These justifications can be shared among different agents in a MAS. 


\subsection{Conceptual Solution}
\label{sec: Conceptual Solution}



Nodes must be able to maintain the integrity of states of management data during the operation of a management CDN, despite message exchange with other nodes. This similarity indicates the use of multi-agent TMS in management CDNs as an interesting possibility. In this Subsection, a conceptual solution is proposed to allow that beliefs can be shared by different nodes (which play the agent role) in a management CDN (which aggregates some MAS characteristics) through \emph{justifications}. In contrast to previous initiatives, our solution focuses on providing a distributed way to maintain the consistency of state of management data by using multi-agent truth maintenance. Justifications improve the alignment of nodes with system-wide objectives (\textit{i.e.}, objectives of the management CDN). Besides, the structure of beliefs and their justifications is a CDN itself.


The datum and its list of possible justifications must be provided by network human operators or expert systems for the management CDN. The associated states of a datum are ``in'' (believed) or ``out'' (disbelieved), according to the presence of its justifications. A datum is labeled ``out'' when it lacks at least one of its associated justifications. We assume that there are not contradictions, so, a datum is labeled ``out'' only if it has not any of its associated justifications.

In the seminal work from Doyle \cite{TMS-Doyle-1979}, a TMS does not require a list of possible justifications for a datum, using an \textit{ad hoc} approach. In order to simplify the implementation and analysis issues, we defined that the datum and its list of necessary justifications must be provided \textit{a priori}. The datum and its list of possible justifications must be provided by network human operators or expert systems. The presence of justifications are based in management information processed by nodes of the management CDN.


For instance, the activation (belief) of a QoS policy (datum) can be justified by a network human administrator command (justification) and an asynchronous signal from a managed device (justification). Listing \ref{lst:example1} shows a possible internal representation of this datum and its justifications using ISO \emph{Prolog} standard \cite{PROLOG-ISO-1995}. In the example, justifications and their presences are represented by facts (Prolog notation) and the datum is represented by a rule (Prolog notation). The last line of Listing \ref{lst:example1} defines that if the justifications ``adm\_cmd'' and ``async\_sig'' are present, the datum ``qos\_pol'' is believed.

\vspace{5pt}

\begin{lstlisting}[caption=Activation of a QoS policy, label=lst:example1]
    justification(adm_cmd).
    justification(async_sig).
    justificationIsPresent(X) :- generated(X).
    justificationIsPresent(X) :- received(X).
    datumIsInternal(qos_pol) :-
     generated(adm_cmd),
     generated(async_sig).
    datum(qos_pol) :-
     justificationIsPresent(adm_cmd),
     justificationIsPresent(async_sig).
\end{lstlisting}

Justifications can be generated by processes inside the node or received through the CDN communication services. Thus, the ``in'' state can assume two additional states: ``internal'', where the datum has only valid internal justifications, and ``external'', where the datum has some valid external justification (provided by other node) \cite{TMS-Huhns-1991}. These additional states are enabled by ``justificationIsPresent'' rules as shown in Listing \ref{lst:example1}. Thus, the presence of a justification is inferred by ``generated'' or ``received'' facts.


The current state of a datum can be checked during the operation of the management CDN. This state is produced through the presence of justifications and their sources. Listing \ref{lst:example2} shows a possible response using the example described in Listing \ref{lst:example1} (activation of a QoS policy). This response denotes that the ``qos\_pol'' datum is believed (\textit{i.e.}, ``:in''). The absence of any of the justifications would change the state of the datum to disbelieved (\textit{i.e.}, ``:out'').

\vspace{5pt}

\begin{lstlisting}[caption=A response from consistency maintenance module, label=lst:example2]
    qos_pol:internal (adm_cmd:mod async_sig:mod)
\end{lstlisting}


The response shown in Listing \ref{lst:example2} also indicates the option ``internal'' being checked. This response shows that the presence of all the justifications was internally generated (indicated by ``:mod'' in Listing \ref{lst:example2}). If the presence of some justification was inferred through belief exchange from other nodes (which would be indicated by ``:msg''), the option ``external'' would be checked. 


Justifications can be also used in an hierarchical way, thus a datum can be used as a justification for other data. Thus, the presence of this kind of justification is controlled by the state of the datum, in other words, if the datum is ``in'' it appears as \emph{present} and if the datum is ``out'' it appears as \emph{absent}. This feature provides support for the representation of more complex data.


The example shown in Listing \ref{lst:example1}, the activation of a QoS policy, can be modified to show the utilization fo hierarchical justifications. We extend the justification ``adm\_cmd'' (``\textit{Administrator Command}'') in order to embrace 2 different commands from a human administrator. Thus, we define a new datum ``adm\_cmd'' and these 2 different commands are represented by 2 new justifications ``adm\_cmd1'' and ``adm\_cmd2'', according to Listing \ref{lst:example3}

\vspace{5pt}

\begin{lstlisting}[caption=A new ``Administrator Command' version, label=lst:example3]
    justification(adm_cmd1).
    justification(adm_cmd2).
    justificationIsPresent(X) :- generated(X).
    justificationIsPresent(X) :- received(X).
    datum(adm_cmd) :-
     justificationIsPresent(adm_cmd1),
     justificationIsPresent(adm_cmd2).
    datumIsInternal(adm_cmd) :-
     generated(adm_cmd1),
     generated(adm_cmd2).
\end{lstlisting}

Modifications are also necessary in ``QoS Police'' datum (qos\_pol). Listing \ref{lst:example4} shows these modifications. The new ``Administrator Command'' datum (adm\_cmd) is used as a justification for the ``QoS Police'' datum through a modification in the ``datum(qos\_pol)'' rule. Besides, the ``datumIsInternal(qos\_pol)'' rule is also modified to allow the option ``internal'' to be properly marked. Thus, if every justification of a datum is internally generated, the datum is processed also as an internally generated justification.

\vspace{5pt}

\begin{lstlisting}[caption=A new ``QoS Police'' version, label=lst:example4]
    datum(qos_pol) :-
     datum(adm_cmd),
     justificationIsPresent(async_sig).
    datumIsInternal(qos_pol) :-
     datumIsInternal(adm_cmd),
     generated(async_sig).
\end{lstlisting}


The conceptual solution for justifications described in this article does not include the representation of justifications that identify belief absences. Traditional TMSs operate with 2 lists: the `` IN'' list, which depicts the accredited beliefs, and the `` OUT'' list, where are represented the discredited beliefs \cite{TMS-Doyle-1979}. Thus, the representation of an absence is defined as part of the ``OUT'' list. In this proposal, there is only a list of justifications, which it is related to accredited beliefs (``IN'' list). However, the definition of a justification that identifies an absence can be performed by the human administrator, using, indirectly, the meaning of the justification. Thus, for example, the absence of a command from an administrator could be described as an ``Administrator Command Off'' fact.


\section{Architecture of management nodes}
\label{sec: Architecture}




Management nodes can be viewed as a \emph{software container} for one or more \emph{management service modules}. These modules perform regular management tasks (\textit{e.g.}, collecting statistics) in each management node, and, in these tasks, modules produce management information. Furthermore, these nodes can appear and disappear in normal operation, behavior that conventional network management systems do not expect from their constitutive elements.


We introduce the \emph{consistency maintenance module} to register and handle the set of beliefs about management data in each management node. This module works associating management data and their respective justifications. In order to simplify the implementation and analysis issues, we define that the consistency maintenance module must receive the datum and its list of necessary justifications (structure).



\begin{figure}[htb]
  \begin{center}
     \includegraphics[scale=.35]{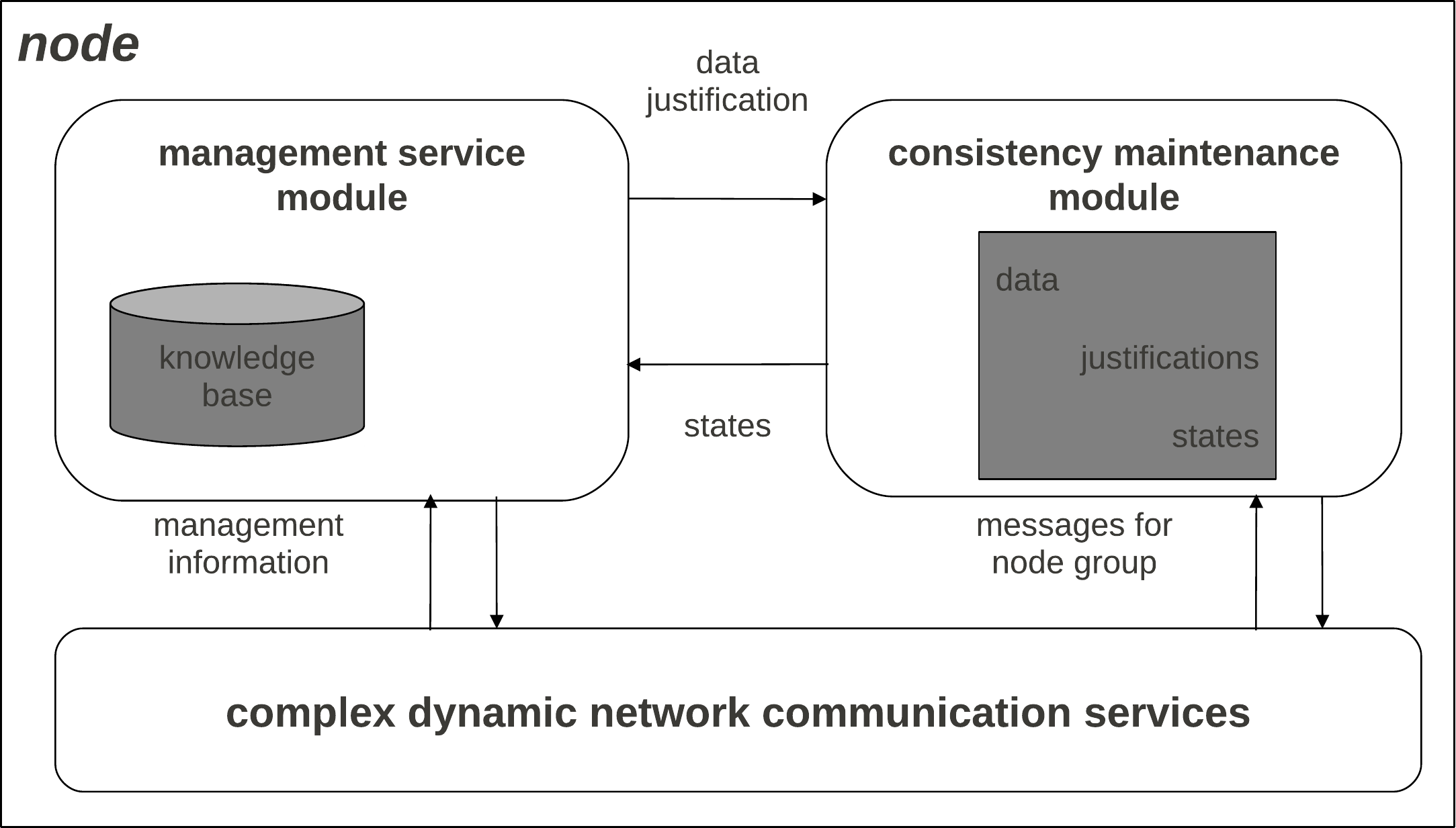}
  \end{center}
  \caption{Architecture of a management node}
  \label{fig:architecture}
\end{figure}

When there is a belief change (\textit{i.e.}, a justification change), the consistency maintenance module uses the CDN communication services to spread the change. Figure \ref{fig:architecture} shows the relation between the consistency maintenance module, management service modules, and CDN communication services.

Beliefs can be represented using different languages. \textit{A priori}, the consistency maintenance module does not require a particular language for belief representation. However, the internal representation of management data and justifications must be unique among the management nodes. In Section \ref{sec: Justification}, we present a representation using facts and rules, concepts from ISO Prolog language.

It is important to warn that the responsibility of logical consistency of a datum and its justifications is not in charge of the consistency maintenance module. The consistency maintenance module verifies whether the defined justifications for a certain datum are present, and, using this information, controls the state of this datum. An advantage of this approach is that the consistency maintenance module is only triggered when there is a change in belief status of shared data, which means it does not introduce overhead for regular management tasks.


The management service modules should inform the consistency maintenance module about their internal beliefs in respect to management data. When the presence of a justification is modified (internally generated or received through a belief exchange message) the consistency maintenance module performs the following steps: unlabels the management datum, includes (or removes) the presence of justification and labels the datum again according to new restrictions. The CDN communication services are used to spread changes, which can change beliefs of other management nodes.

The management service modules are responsible for querying and requiring services from the consistency maintenance module, according the demands of their processes. Some similar services employ different strategies such as ``\textit{publish/subscribe}'' schemes \cite{AUTONOMIC-Vanrenesse-2003} or ``\textit{watches} (changes trigger the transmission of a packet) \cite{AUTONOMIC-Zookeeper-2008}. These strategies could be implemented in the proposed consistency maintenance module, however, initially we keep this operation in charge of management service modules.


It is important to stress that there is only one consistency maintenance module inside a management node, thus, it is not specific to a management service module. Therefore, every management service module in this node interacts with the same consistency maintenance module. This fact can be explored for the integration of different management services. For instance, a policy processing module, a fault handling module, and a configuration management module (possibly using different languages for representing management data) could be integrated by the consistency maintenance module through justifications.


The services offered by management service modules and the consistency maintenance module may be locally available through the use of interprocess communication. These services can be offered in different ways, such as, for example, using a message bus like \emph{D-Bus} (\textit{Desktop Bus}) \cite{DBUS-Pennington-2010} or through interprocess communication protocol like \emph{DCOP} (\textit{Desktop COmmunication Protocol} \cite{DCOP-Moore-2010}.


The consistency maintenance module is conceived considering that there is support for group organization (\textit{i.e.}, node groups) through management services modules. Thus, nodes that  have a specific management service module are organized into a group (without human intervention) and nodes can participate of several groups accordingly to modules that they have \cite{AC_NM-Marquezan-2007}. In this context, the management CDN must support group communication services. These services are often supported in overlay networks. For instance, the ``PeerGroup Service'' from JXTA \textit{framework} \cite{JXTA-Gong-2001} is an implementation for an partially structured overlay. An interesting possibility to support group communication services is the utilization of \emph{multicast}, but this utilization is not a \textit{sine qua non} condition.


The consistency maintenance module offers an open environment for rapid and dynamic resource integration of management service modules are formed with no central authority. This is feasible because each node runs its own consistency maintenance module, which keeps its own data structure and handles belief exchange through messages among other nodes. These groupings have analogous characteristics with \emph{federations}, a concept brought from MAS literature \cite{MAS-Horling-2005}.


The consistency maintenance module is offered through a simple interface that management service modules must use. The number of supported operations is restricted to ease the implementation in management service modules. Besides, communication details are not specified in supported operations, so changes in communication strategies do not lead to adaptations in the interfaces of management service modules. 





\section{Case studies}
\label{sec: Case Study}



Case studies can be used to illustrate the utilization of justification for management data through the consistency maintenance module in a management CDN. Network management, in which we aimed at the present proposal, is an interesting context for CDNs because of their desirable characteristics. For example, one of this characteristics is that some CDNs display a remarkable degree of tolerance against errors \cite{CDN-Albert-2000}. However, the consistency of state of management data can impose challenges to management CDNs (as discussed in Section \ref{sec: Scenario}).

%



In this Section, 2 case studies are presented. In the first subsection, the consistency maintenance of state of obligation policies in the form of Event-Condition-Action rules is described. In the following subsection, collaborative fault management of links in access networks through failure notification sent by devices and human knowledge about these notifications is characterized. 

\subsection{Consistency maintenance of state of obligation policies}

The first case study presented is an illustration of the consistency maintenance of state of policies in a management CDN (a simpler and shorter version of this case study was described in a previous work \cite{ANM-Nobre-2010}). We use obligation policies in the form of Event-Condition-Action (ECA) rules as our model. Using the concepts from our proposal, policy state (\textit{i.e.}, policy activation) is a management datum, and event and conditions are justifications in the consistency maintenance module. This module provides an integration point to enable a coherent global behavior of policy-driven (or -enabled) management CDNs. 


We describe the policy used in this case study using \emph{Ponder2} \cite{POLICY-Ponder2-2009}. Ponder2 is a toolkit that supports the specification and enforcement of policies in the form of Event-Condition-Action (ECA) rules. When the occurrence of an event is announced to Ponder2, the obligation policy interpreter matches the notification against the registered policy events, hands the event to the relevant policies, which then evaluate their condition(s) and if they succeed, invoke the action(s) specified in the policy.

The Listing \ref{lst:case1} shows the XML encoding of an obligation policy that will respond to events of type \emph{/event/humanResourcesProcedureEvent}. In the example, Ponder2 checks whether the load on the router R1 is low, and the current time is later than 18:00. In this example, if the conditions are satisfied, the policy invokes an action on the router R1 to reserve 10\% of bandwidth in ``QoS1'' profile.

\vspace{5pt}

\begin{lstlisting}[language=XML, caption=An obligation policy, label=lst:case1]
<use name="/policy">
 <add name="AdjustQoSPolicy">
  <use name="/template/policy">
   <create type="obligation" 
   event= "/event/humanResourcesProcedureEvent" 
   active="true">
     <arg name="R1_load"/>
     <arg name="daytime"/>
     <condition>
      <and>
       <eq>!R1_load;<!-- -->low</eq>
       <gt>!daytime;<!-- -->18:00</gt>
      </and>
     </condition>
     <action>
      <use name="/routers/R1">
       <modify profile="QoS1" value="10%"/>
      </use>
     </action>
   </create>
  </use>
 </add>
</use>
\end{lstlisting}


To maintain consistent the state of this policy, the consistency maintenance module receives the datum ``\textit{Adjust QoS Policy}'' (adj\_qos\_pol) and its justification list, composed by ``\textit{human resources procedure event received}'' (hr\_proc\_evt), ``\textit{R1 Low Load Matched}'' (R1\_load\_mat), and ``\textit{Daytime Matched}'' (dt\_mat). The Listing \ref{lst:case2} shows the internal representation of this datum and its justifications.

\vspace{5pt}

\begin{lstlisting}[caption=Obligation policy internal representation, label=lst:case2]
    justification(hr_proc_evt).    
    justification(R1_load_mat).
    justification(dt_mat).
    justificationIsPresent(X) :- generated(X).
    justificationIsPresent(X) :- received(X).
    datum(adj_qos_pol) :- 
     justificationIsPresent(hr_proc_evt),
     justificationIsPresent(R1_load_mat),
     justificationIsPresent(dt_mat).
    datumIsInternal(adj_qos_pol) :-
     generated(hr_proc_evt),
     generated(R1_load_mat),
     generated(dt_mat).
\end{lstlisting}


We will present a situation as an example of use of our proposal: due to lack of synchronization, some peers disagree about a temporal condition. This situation could lead to to an inconsistent state of node group. In our proposal, after changing the evaluation of the temporal condition, policy processing component informs the belief change to the consistency maintenance module. If the new belief implies a justification change, a message will be sent to peer group informing the change. When receiving this message, the consistency maintenance module of other nodes checks whether the received justification change is consistent with their knowledge base. If it is not, the consistency maintenance module updates the justification and verifies if this update implies in other changes. The Listing \ref{lst:case3} shows the answer from the consistency maintenance module in this situation.

\vspace{5pt}

\begin{lstlisting}[caption=A response for the ``Adjust QoS Policy'' datum, label=lst:case3]
adj_qos_pol:external (hr_proc_evt:mod R1_load_mat:mod dt_mat:msg)
\end{lstlisting}


The response produced by the consistency maintenance module indicate that the state of the ``\textit{Adjust QoS Policy}'' (adj\_qos\_pol) datum is ``in'' and the option``external'' is marked. Besides, the response also shows that the presence of the ``\textit{Daytime Matched}'' (dt\_mat) justification was received as a belief change message (indicate by ``:msg'' in the Listing\ref{lst:case3}). Thus, the state of the management datum is consistent in node group and the management CDN presents a coherent behavior

The consistency maintenance of state of policies in management CDNs is traditionally performed through centralized entities, such as external repositories \cite{ANM-Marquezan-2008}. In P2P-based ANM systems, this centralization is normally performed through the use of ``super peers'' \cite{ANM-Kamienski-2008} \cite{ANM-Fallon-2007}. The utilization of a centralized approach brings concerns in scalability and robustness and imposes difficulties in the integration of different information sources. 


It is also possible to use a datum as a justification for another datum (according to Section \ref{sec: Justification}). In our example, we can extend the justification ``\textit{R1 Low Load Matched}'' to include different ``loads'' of the router R1, such as memory and CPU load. Thus, we define the datum ``\textit{R1 Low Load Matched}'' (R1\_load\_mat) and its justification list, composed by ``\textit{CPU Low Load Matched}'' (cpu\_load\_mat) and ``\textit{Memory Low Load Matched}'' (mem\_load\_mat). The Listing \ref{lst:case4} shows the representation of this datum and its justifications. 

\vspace{5pt}

\begin{lstlisting}[caption=A new ``R1 Low Load Mat'' datum, label=lst:case4]
  justification(cpu_load_mat).    
  justification(mem_load_mat).
  justificationIsPresent(X) :- generated(X).
  justificationIsPresent(X) :- received(X).
  datum(R1_load_mat) :- 
   justificationIsPresent(cpu_load_mat),
   justificationIsPresent(mem_load_mat).
  datumIsInternal(R1_load_mat) :-
   generated(cpu_load_mat),
   generated(mem_load_mat).
\end{lstlisting}


It is also required a little change in the representation of the ``\textit{Adjust QoS Policy}'' datum. The rule that defines if the option ``internal'' is marked must be changed as well (\emph{datumIsInternal (adj\_qos\_pol)}). The Listing \ref{lst:case5} shows the new version of ``Adjust QoS Policy'' datum. The feature presented in the last example, the support for hierarchical justifications, enhances the representation of complex data. 

\vspace{5pt}

\begin{lstlisting}[caption=A new ``Adjust QoS policy'' version, label=lst:case5]
  datum(adj_qos_pol) :- 
   justificationIsPresent(hr_proc_evt),
   datum(R1_load_mat),
   justificationIsPresent(dt_mat).
  datumIsInternal(adj_qos_pol) :-
   generated(hr_proc_evt),
   datumIsInternal(R1_load_mat),
   generated(dt_mat).
\end{lstlisting}


Justifications can be used to provide explanations about specific data for users, such as policies \cite{TMS-Kagal-2008}. In the network management context, these explanations can be used by human network administrators in order to improve the execution of management tasks and the user comprehension about these tasks. Thus, the utilization of justifications can increase confidence in the results of management tasks (\textit{e.g.}, in policy enforcement) \cite{TMS-Kagal-2008}. 

\subsection{Collaborative fault management of links in access networks}
\label{subsec: ethernet}

The second case study presented is an illustration of the collaborative fault management of interdomain links (\textit{e.g.}, service provider and consumer) through failure notification sent by devices and human knowledge about these notifications (a simpler and shorter version of this case study was described in a previous work \cite{ANM_ETH-Nobre-2009}). The integration of these information (failure notification in addition to human knowledge) produces a management data, which can assume different states. For example, these data can be used against Service-Level Agreements (SLAs) to support or clarify service level claims.


Among access network technologies in metropolitan networks, Ethernet is one of most interesting and promising choice, thus, we choose this technology to build our case study. In this context, an access network link is an \emph{Ethernet Virtual Connection} (EVC) \cite{METRO_ETH-McFarland-2005}. Fault management in this link is done through \emph{Alarm Indication Signal} (AIS) messages \cite{METRO_ETH-McFarland-2005}. These messages are triggered when a link failure occurs. Thus, AIS messages provide asynchronous notification to other elements in the network that there is a fault in the Ethernet network \cite{METRO_ETH-Sanchez-2008}. The efforts to manage layer 2 Ethernet services must consider an overlayed IP infrastructure \cite{METRO_ETH-Ryoo-2008}.


In the present case study, nodes have a management service module that collects AIS messages and another module that collects information from human administrators. Besides, there is the consistency maintenance module, responsible to integrate the information from both management service modules and maintain the consistency of the state of management data.

The consistency maintenance module receives the ``\textit{Link Fault Detected}'' (link\_flt\_det) datum and its list of justifications, composed by ``\textit{Service Operator Detection}'' (srv\_prv\_det), ``\textit{Service Consumer Detection}'' (svr\_con\_det), and ``\textit{Device Notification Received}'' (dev\_not\_rcv). The presence of these justifications is provided by management service modules and kept inside the node group that offers this management service. The Listing \ref{lst:case6} shows the representation of this datum and its justifications.

\vspace{5pt}

\begin{lstlisting}[caption=``Link Fault Detected'' datum, label=lst:case6]
  justification(srv_prv_det).    
  justification(srv_con_det).
  justification(dev_not_rcv).
  justificationIsPresent(X) :- generated(X).
  justificationIsPresent(X) :- received(X).
  datum(link_flt_det) :- 
   justificationIsPresent(srv_prv_det),
   justificationIsPresent(srv_con_det),
   justificationIsPresent(dev_not_rcv).
  datumIsInternal(link_flt_det) :-
   generated(srv_prv_det),
   generated(srv_con_det),
   generated(dev_not_rcv).
\end{lstlisting}

The presence of justifications can be defined by the management service modules or the consistency maintenance module. The ``justificationIsPresent'' rules represent the possibility to define the presence of justifications through ``generated'' facts, when the presence is internally generated by the management service modules , and `received'' facts, when the presence is received through belief change message from other management nodes (using the consistency maintenance module).


We will present a situation as an example of use of our proposal: due to problem in the reception of an AIS message (represented by the ``Device Notification Received'' justification), some nodes of the node group disagree about the fault occurrence in a network link (\textit{i.e.}, an EVC). Despite this problem, commands performed by human network administrators in both domains acknowledge the fault and these are received by all the nodes of the node group as the respective justifications (``\textit{Service Operator Detection}'' and ``\textit{Service Consumer Detection}''). This situation can become the state of the node group inconsistent. However, belief exchange can correct this inconsistency. The Listing \ref{lst:case7} shows the answer from the consistency maintenance module in this situation.

\vspace{5pt}

\begin{lstlisting}[caption=A response for a ``Link Fault Detected'' datum, label=lst:case7]
link_flt_det:external (srv_prv_det:mod srv_con_det:mod dev_not_rcv:msg)
\end{lstlisting}


The response produced by consistency maintenance module indicates that the state of the ``Link Fault Detected'' is ``in'' and the option ``external'' is marked. Besides, it can be verified that the presence of the ``Device Notification Received'' was received through belief change messages (indicated by ``:msg'' in Listing \ref{lst:case7}). Thus, the state of this management datum is maintained consistent within the node group and the management CDN presents a coherent behavior.


The \textit{Operations, administration, and maintenance} (OAM) tools of Ethernet enable the use of a signaling intended to differentiate a fault condition and an intentional administrative block in the EVC (\textit{e.g.}, for diagnostic purposes). This signaling is performed by the ``\textit{Locked Signal Function}'' (LCK) message \cite{METRO_ETH-Ryoo-2008}. In this context, the link fault has to be detected when the LCK message is not perceived. According to the conceptual solution described in Section \ref{sec: Justification}, the definition of a justification that identifies an absence can address the belief about LCK messages. 


The justification list of ``Link Fault Detected'' datum must be modified to introduce a new justitication, ``\textit{Device Diagnostics Off}'' (dev\_dia\_off). This justification defines that the LCK is absent, therefore the managed EVC is not administrative blocked. The Listing \ref{lst:case8} shows a new version of the ``Link Fault Detected'' datum and its justifications. 

\begin{lstlisting}[caption=A new ``Link Fault Detected'' version, label=lst:case8]
  datum(link_flt_det) :- 
   justificationIsPresent(srv_prv_det),
   justificationIsPresent(srv_con_det),
   justificationIsPresent(dev_not_rcv),
   justificationIsPresent(dev_dia_off).
  datumIsInternal(link_flt_det) :-
   generated(srv_prv_det),
   generated(srv_con_det),
   generated(dev_not_rcv)
   generated(dev_dia_off).
\end{lstlisting}


Fault management of access links is a management task traditionally performed through stand-alone centralized systems. Usually, these systems are a \emph{network management system} (NMS) (which collects devices notifications); and a \emph{trouble ticket system} (TTS) (which collects information from human network administrators). The traditional procedure brings concerns in scalability and robustness due to the centralization, since each one of these systems is a single point of failure (SPoF) and a bottleneck.


The traditional procedure for fault management of network links also imposes difficulties in the integration of the management information from different sources (NMS and TTS). These systems may operate internally with different data models and languages.


\section{Communication strategies}
\label{sec: Communication}


The consistency maintenance module handles the message exchange through CDN communication services. In this process, belief exchange requests are adapted in messages to be spread in management CDN and vice-versa. The management CDN is modeled as an unstructured overlay network, thus there is no relation between the information stored at an management node and its position in the overlay topology. Besides, the set of nodes can be temporally dynamic.


We use the premise that there is support for group organization (\textit{i.e.}, node groups) through management services modules. Thus, nodes that have a specific management service module are organized into a group (without human intervention) and nodes can participate of several groups accordingly to modules that they have. Aggregations are an interesting abstraction to improve the scalability features in distributed systems \cite{AUTONOMIC-Vanrenesse-2003}.


The exchange of beliefs about management data is done asynchronously and we do not consider the message exchange to be reliable. It is well known that the utilization of asynchronous unreliable distributed systems imposes challenges to achieve consistency in shared data \cite{FT-Fischer-1985}. Thus, the consistency model used is non-deterministic, in other words, it uses a ``weak'' notion of consistency. This model is adopted for scalability, robustness, and update dissemination issues. Given a belief ``X'' that depends on some other belief ``Y'', when an update is made to ``Y'', it is eventually reflected in ``X''. To achieve consistency, all belief changes about a specific datum must be propagated to all nodes that share this datum. Some authors call a similar notion as ``eventual consistency'' \cite{AUTONOMIC-Vanrenesse-2003}.

The methods used for message exchange inside the AMD are modeled using concepts from biology-inspired distributed computing models \cite{BIO-Babaoglu-2006}. Among these models, proliferation-based ones are an interesting choice for communication requirements of our proposal. Several proliferation-based models are described in distributed systems literature. In the present proposal, a belief change possibly undergoes proliferation at nodes visited, where each node calculates the probability of forwarding the belief change using a proliferation controlling function. All nodes in the node group run exactly the same proliferation controlling function and proliferation can be initiated from any node in the node group. We have chosen \emph{replication} as the initial proliferation mechanism in node groups. This mechanism can support several communication strategies. Processes based on replication replication are commonplace in Nature (\textit{e.g.}, epidemic spreading, mammalian immune system) \cite{BIO-Babaoglu-2006}.


The communication strategies described in the following Subsections fit with consistency model presented in this Section. Belief computations could occur concurrently and changes in beliefs are replicated possibly among all nodes within a node group. Different nodes in a node group are not guaranteed to have identical copies of current set of beliefs even if queried at the same time, and not all nodes are guaranteed to perceive each and every update to a current set of beliefs.


In this Section, we present 2 communication strategies developed in the context of the present investigation. Initially, the Unbridled replication is presented in the first Subsection. Afterwards, the controlled replication is present in the second Subsection.

\subsection{Unbridled replication}


In unbridled replication, management nodes spread messages to replicate every change in justifications among the participating entities (\textit{i.e.}, nodes of a specific node group) \cite{ANM_ETH-Nobre-2009}. This mechanism is restricted to node group, fulfilling the criterion of robustness for small node groups. Besides, belief changes are replicated as soon as they occur.


The unbridled replication is implemented through the flooding of belief changes inside node groups. For example, flooding techniques have been usually used to implement search operations in unstructured networks. Flooding fulfills the criterion of robustness and also gives fast results in operations inside node groups. However, it produces a huge number of messages which increases the bandwidth used to carry network management traffic and the amount of computational resources necessary to message processing.


The scalability of the unbridled replication is directly related to the definition of node groups, since this operation (message exchange) is restricted to each node group (similarly to \emph{domain}, a concept used in \textit{Astrolabe} \cite{AUTONOMIC-Vanrenesse-2003}). The number of messages within the node group is controlled only through the discard of already received messages. Thus, a node can spread a belief change message, despite having received several messages informing about the same belief change.

\subsection{Controlled replication}


The \emph{controlled replication} was developed after preliminary investigations regarding the limitations of the unbridled replication for large node groups \cite{ANM-Nobre-2010}. Our goal is to use a much lower number of messages in belief changes (than in unbridled replication). Controlled replication forwarding means some belief changes will not be forward due to restriction rules. The idea behind this control is that this way we can minimize redundant network utilization.


The replication is controlled by a \emph{Replication Controlling Function} (RCF). The RCF defines the probability '\textit{P}' of a belief change to be forwarded to the node group. Broadly speaking, this probability could be defined using different parameters of the CDN operation. Initially, we define the RCF as shown in Equation \ref{eq:1}:


\begin{equation}
\label{eq:1}
P(\rho, \eta b  ) = \frac{\rho}{1 +\eta b}
\end{equation}


where $\eta$\textit{b} represents the number of received messages of a specific belief change within a ''backoff delay`` (\textit{T}) and $\rho$ is the positive proliferation constant. The essence of the function is that proliferation of a belief change should decrease with the reception of multiple copies of the same belief change (message). The equation \ref{eq:1} is evaluated at time \textit{T}.


When the belief change is produced internally, the node sends belief exchange with \textit{P} = 1 (i.e., the belief change is always sent) to the node group. When the belief change is received by CDN communication services, the reception of new messages informing the belief change already received during the backoff delay decreases the probability of this belief change to be forwarded to the node group. Note that when \textit{T} = 0 or $\eta$\textit{b} = 0, the controlled replication has the same behavior of unbridled replication.

\section{Simulation Experiments}
\label{sec: Evaluation}


Scalability and robustness are some of the most important motivations for using decentralization in the infrastructure of different systems \cite{AUTONOMIC-Mccann-2004} \cite{BIO-Babaoglu-2006}. As previously stated, we expect that the introduction of multi-agent truth maintenance features keeps desired properties of a management CDN, maintaining each node as an independent and self-sustainable entity. As many systems have demonstrated, a system that does not share resources can scale almost infinitely simply by adding constitutive elements (\textit{e.g}, nodes in a management CDN). Besides, maintaining the independence of each node, single points of failure (SPoF) are eliminated.

The evaluation of our proposal can be performed in different ways. To enable a fully controlled environment for the evaluation, we evaluated the system with simulation experiments. A management CDN is modeled as a P2P-based ANM system, thus, the nodes are simulated as peers and node groups as peer groups. In these experiments, we present simulation results that support our scalability and robustness claims. 

 
The simulation experiments were implemented in Java using \emph{PeerSim} \cite{P2P_SIM-Montresor-2009}, an open source event-based simulator of P2P systems. The system version used has the ability to simulate failures in peers and message exchange, and the overlay is built randomly. The experiments use a simple model of transport layer that can emulate some characteristics, such as loss and delay probabilities. In addition, a peer is chosen randomly as the primary source of changes to not affect measurements and message delay is controlled. All peers in peer groups run exactly the same communication strategy. In spite of being event-driven simulations, events are evaluated in cycles of the simulator. We simulate up to 100,000 peers in a peer group.


In this Section, we present results of simulation experiments performed with the communication strategies described in Section \ref{sec: Communication}. First, experiments using unbridled replication in order to provide belief exchange are presented. Then, experiments using controlled replication are also presented.

\subsection{Unbridled replication}


Unbridled replication is evaluated through 2 simulation experiments. In these experiments, we varied the number of peers of the peer group from 4 to 14 (we do not expect large peer groups using unbridled replication). Each experiment was conducted at least 10 times. In the experiments, the variance observed was low.

\begin{figure}[htb]
  \begin{center}
     \includegraphics[scale=.85]{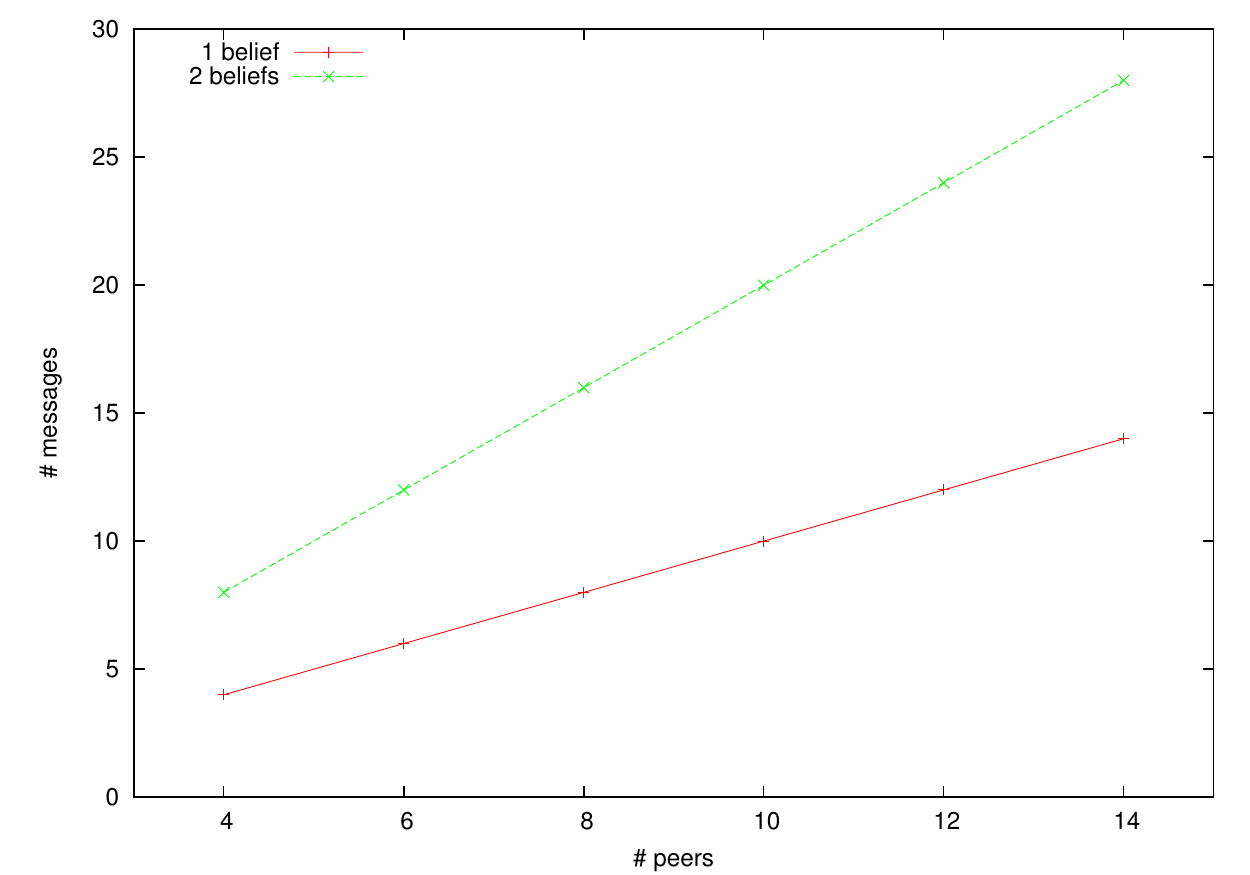}
  \end{center}
  \caption{Message exchange due to a belief change - Unbridled Replication} 
  \label{fig:graph_sim1}
\end{figure}

In the first experiment, it is measured the number of messages exchanged to spread belief changes in the peer group. This number must be considered as an important cost of the peer group operation, thus, it is important for scalability analysis. Besides, we consider the number of transmitted messages as indicative of network load. In this experiment there were no faults in peers or in message exchange. We show the results in Figure \ref{fig:graph_sim1}.

Our proposal shows acceptable scalability characteristics on number of exchanged messages, since this operation (belief exchange) is restricted to each peer group. The experiment shows that our system behaves like we expected, without stability and convergence problems. Network load grows linearly with the number of participating peers, thus we can infer the behavior trend of peer groups with larger number of participating peers. Of course, an efficient operation of large peer groups needs modifications in the communication strategy, such as using a mechanism to control the replication.

\begin{figure}[htb]
  \begin{center}
     \includegraphics[scale=.85]{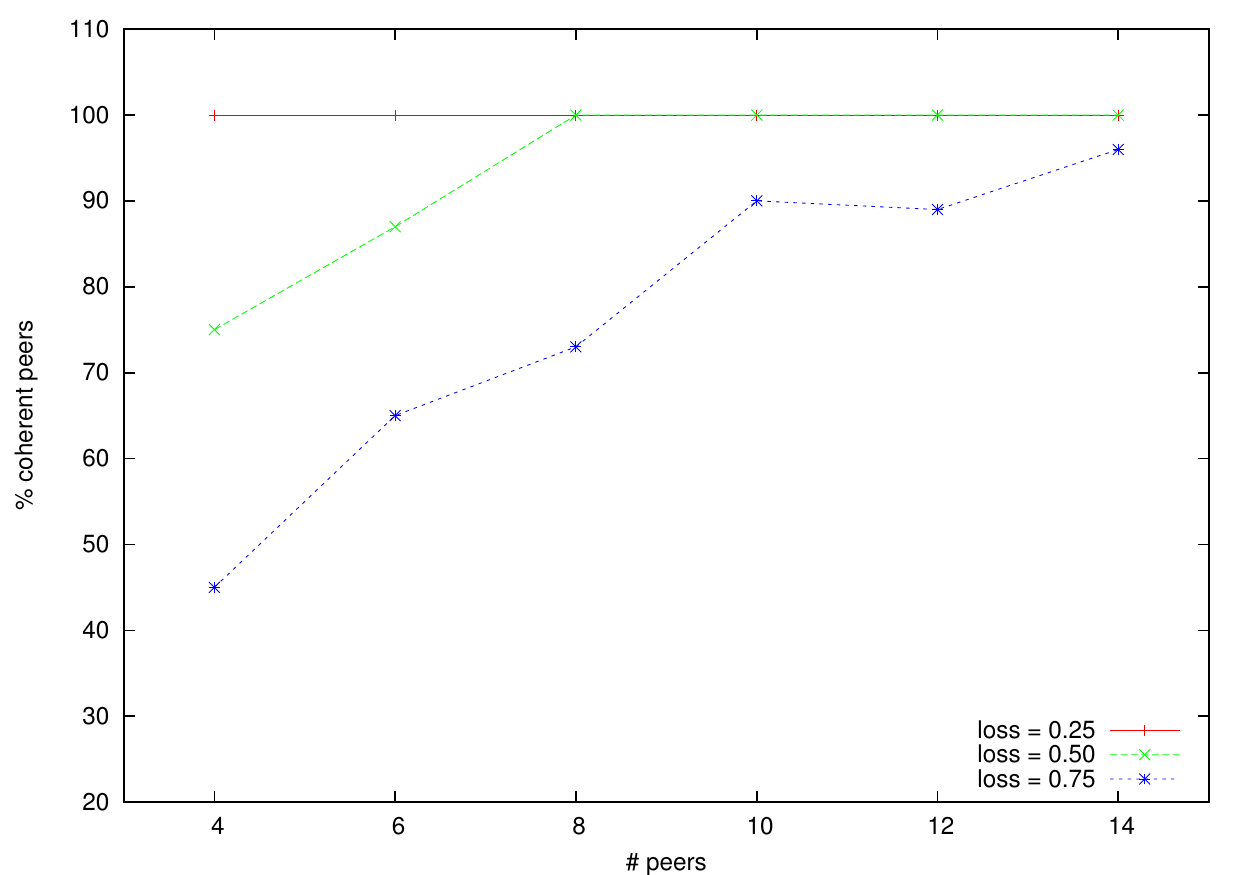}
  \end{center}
  \caption{Coherent peers after a belief change - Unbridled Replication} 
  \label{fig:graph_sim2}
\end{figure}

In the second experiment, we determined the influence of message loss on the dissemination of a justification change. In this experiment, we varied the message loss probability with following values: 25\%, 50\%, and 75\% (respectively, 0.25, 0.5, and 0.75 as indicated in Figure \ref{fig:graph_sim2}). Using the case study described in Subsection \ref{subsec: ethernet}, we would probably observe such message loss (specially 75\%) due to faulty or overloaded network devices (\textit{e.g.}, ethernet interfaces) and/or network links (\textit{e.g.}, EVCs). Since this case study is aimed at fault management (considering an overlayed IP infrastructure), our proposal must behave acceptably even in bad network conditions. In Figure \ref{fig:graph_sim2}, we show the average percentage of coherent (and correct) peers after message exchange to cease.

The experiment shows the influence of message loss in the replication process. As can be seen from the results in Figure \ref{fig:graph_sim2}, high loss probabilities do lead to less consistency in peer group, but, even with a few participating peers, the percentage of coherent peers is substantial. Besides, more participating peers in peer group decrease the influence of loss probability.

The results show some fault-tolerance features, since the peer group operation is not highly sensitive to peer crashes and message losses. But an increase in number of peers also leads to an important increase in the number of exchanged messages, so the robustness advantages come at some cost. Aggregation of peers is a fundamental abstraction for scalability using unbridled replication as the communication strategy. Messages are exchanged only within the peer group, and a high number of peers in a peer group is not expected when this strategy is chosen. 

\subsection{Controlled replication}

Controlled replication is evaluated through 2 simulation experiments. In these experiments, we varied the number of peers of the peer group from 5 to 1000. The parameters used in the Replication Controlling Function (RCF) are informed in each experiment. In the experiments, the variance observed was low.

\begin{figure}[htb]
  \begin{center}
     \includegraphics[scale=.85]{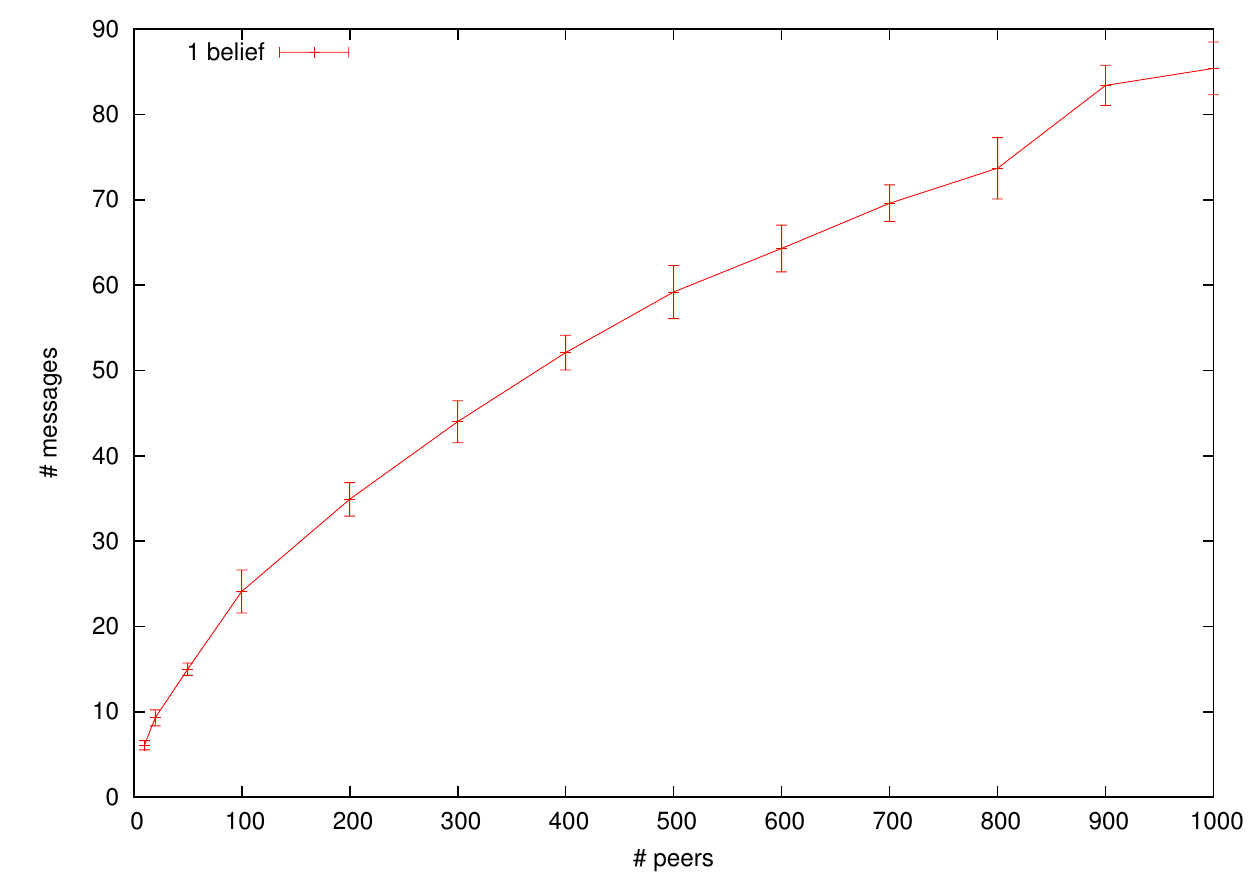}
  \end{center}
  \caption{Message exchange due to a belief change - Controlled replication}
  \label{fig:graph_sim3}
\end{figure}

In the first experiment ($\rho$ = 1, \textit{T} = 3), the number of messages exchanged to spread a justification change in the peer group is measured. This number must be considered as an important cost of the peer group operation, thus, it is important for scalability analysis. Besides, we consider the number of transmitted messages as indicative of network load. In this experiment there were no faults in peers or in message exchange. In Figure \ref{fig:graph_sim3} we report the measured average and 95\% confidence intervals for each peer group size.

The experiment shows that our system behaves as we expected, without stability and convergence problems. The main reason for this is that chosen RCF is cost-effective when covering large peer groups. We consider the number of transmitted messages as indicative of network load. Even though we do not expect this situation in P2P-based ANM systems (and in other management CDNs), it is important to perform this experiment to infer the behavior trend of peer groups with larger number of participating peers.

\begin{figure}[htb]
  \begin{center}
     \includegraphics[scale=.85]{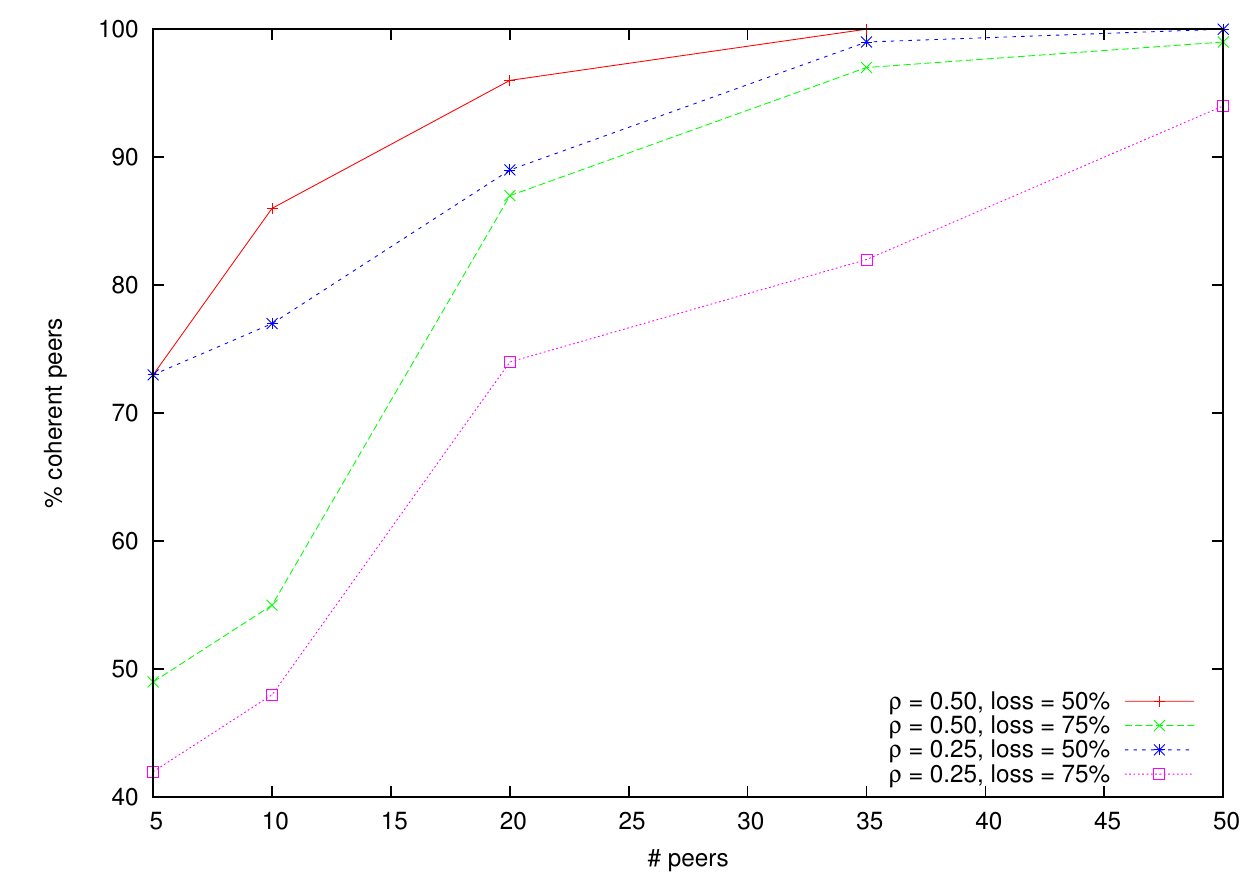}
  \end{center}
  \caption{Coherent peers after a belief change - Controlled replication}
  \label{fig:graph_sim4}
\end{figure}

In the second experiment, we determined the influence of message loss on the dissemination of a justification change. In this experiment, we varied the message loss probability with following values: 50\%, and 75\%. During the experiments, we observed a significantly correlation between message loss and positive proliferation constant ($\rho$) on results, thus, we also varied $\rho$ in this experiment with following values: 0.5 and 0.25. In Figure \ref{fig:graph_sim4}, we show the average percentage of coherent (and correct) peers after message exchange to cease.

The experiment shows the influence of message loss and positive proliferation constant ($\rho$) in the controlled replication scheme. As can be seen from the results in Figure \ref{fig:graph_sim4}, high loss probabilities do lead to less consistency in peer group, but, even with a few participating peers, the percentage of coherent peers is substantial. Besides, more participating peers in peer group and higher values of $\rho$ decrease the influence of loss probability.

\section{Related work}
\label{sec: Related Work}

In recent years, several research efforts on services for consistency of shared information have been carried out distributed systems community. These services can be used as a basic building block for distributed applications. A management CDN can be viewed as a distributed application and it can be improved with characteristics found in these services. In this section, we cover some of the most prominent investigations.

\emph{ZooKeeper} \cite{AUTONOMIC-Zookeeper-2008} is a coordination service for distributed applications. It exposes a simple API that distributed applications can be built upon to implement higher level services for synchronization, data diffusion, and publish-subscribe schemes. \emph{ZooKeeper} use distributed server databases for read operations, however, write operations use a ``leader'' server (\textit{i.e.}, centralized database) to assure the consistency of the database.

\emph{Astrolabe} \cite{AUTONOMIC-Vanrenesse-2003} is a distributed information management service. It works locating and collecting the status of a set of servers and reporting summaries of this information. \emph{Astrolabe} is implemented using a P2P overlay, where every peer run an \emph{Astrolabe} agent (\textit{i.e.}, in a MAS fashion). However, \emph{Astrolabe} was developed primarily using simple data models. Besides, its operation is aimed at read-oriented applications.

\emph{Scalable Distributed Information Management System} (SDIMS) \cite{AUTONOMIC-Yalagandula-2004} is a service to aggregate information about large-scale network systems. The service is built using ideas from \emph{Astrolabe} \cite{AUTONOMIC-Vanrenesse-2003} and Distributed Hash Tables (DHT). However, as in most DHT approaches, consistency and replication issues are a known challenge.

%

The presented efforts show interesting characteristics for consistency of shared information in distributed systems. However, these efforts have vulnerabilities which make them not appropriate for management CDNs, such as centralization \cite{AUTONOMIC-Zookeeper-2008}, simple data models \cite{AUTONOMIC-Vanrenesse-2003}, and replication issues \cite{AUTONOMIC-Yalagandula-2004}.


\section{Final remarks}
\label{sec: Conclusion}


The support of new demands faced by traditional network management is a key research issue in the network management area. The utilization of Complex Dynamic Networks (CDNs) can be a feasible approach for these demands, specially when it is used together with automation features. However, the consistency of state of management imposes challenges for management CDNs.


In this paper we use multi-agent truth maintenance features and dynamic process as communication strategies to improve the consistency of state of management data in management CDNs. Our proposal aims at the integration of data used by the entities that form these systems (\textit{i.e.}, nodes), through the utilization of belief about management data. We have also presented evaluations of this proposal through simulation experiments. In addition, we have described case studies to show the possibilities of our proposal.


Although the proposal shows good results in evaluations performed until the present moment, it is necessary to evaluate more complicated cases, in number of nodes and node groups, and in the participation of an nodes in different node groups. We are also looking at additional settings that could lead to important effects, such as network partitions. Thus, we are currently pursuing new experiments with \emph{PeerSim}.



\bibliographystyle{abbrv}
\bibliography{jcnobre-consistency-full}

\end{document}